\documentclass[aps,amsfonts,pra,preprint,singlecolumn,showpacs]{revtex4}

\usepackage{hyperref}
\usepackage{sidecap}
\usepackage{graphicx}
\usepackage{color}
\usepackage{mathrsfs}
\usepackage{amsmath}
\usepackage{dsfont}

\def\NOT(#1,#2){\OneQubitGate(#1,#2){$X$}}
\begin{document}

\title{Hamiltonian Characterization for Dipolar Coupled Spin Systems}
\author{Denis-Alexandre Trottier$^{1}$, Virginia Jauregui-Villanueva$^{2}$, and Jingfu Zhang$^1$\\
 $^1$Institute for Quantum Computing and Department of Physics,
University of Waterloo, Waterloo, Ontario, Canada N2L 3G1\\
 $^2$ Instituto de Ciencias Nucleares, Universidad Nacional Aut\'onoma de M\'exico, Apdo. Postal 70-543, 04510, M\'exico D.F\\}
%\date{\today}

\begin{abstract}
Measuring the Hamiltonian of dipolar coupled spin systems is usually a difficult task due to the high complexity of their spectra. Currently, molecules with unknown geometrical structure and low symmetry are extremely tedious or impossible to analyze by sheer spectral fitting.
We present a novel method that addresses the problem of spectral analysis, and report experimental results of extracting, by spectral fitting, the parameters of an oriented 6-spin system with very low symmetry in structure, without using  {\it a priori} knowledge or assumptions on the molecular geometry or order parameters. The advantages of our method are achieved with the use of a new spectral analysis algorithm - NAFONS (Non-Assigned Frequency Optimization of NMR Spectra), and by the use of simplified spectra obtained by transition selective pulses.
This new method goes beyond the limit of spectral analysis for dipolar coupled spin systems and is helpful for related fields, such as quantum computation and molecular structure analysis.
\end{abstract}
 \maketitle

\section{Introduction}
Obtaining the Hamiltonian of a system by the extraction of its parameters from experimentally measured data is an inverse problem, one of the fundamental problems in physics. In order to control a system, as in quantum information processing \cite{NielsenChuang}, this task is critically important, because current optimal control algorithms, such as gradient ascent pulse engineering \cite{grape} and strongly modulating pulse \cite{SMP} algorithms,
depend on the full information of the Hamiltonian. Moreover, the Hamiltonian of spin systems provides valuable information for molecular structure analysis \cite{structure}.

In dipolar coupled spin systems, such as molecules dissolved in liquid crystal solvents in Nuclear Magnetic Resonance (NMR) experiments \cite{books}, the Hamiltonian is not naturally diagonal due to the interaction terms with dipolar couplings, which are usually too strong for the weakly coupling approximation to be satisfied. Consequently, the spectra are usually very complex in multiple-spin systems, where the number of peaks corresponding to single coherence increases rapidly with the number of interacting spins. Furthermore, in liquid crystal solvents, the dipolar couplings depend on the solute's size and shape, and are scaled by the order parameters, which are sensitive to multiple factors, such as the characteristics of the solvents, magnetic fields, temperature, etc, making almost impossible the theoretical calculation for obtaining the dipolar couplings. First-order analysis of dipolar coupled spectra are usually not possible and the Hamiltonian has to be diagonalized numerically.

Measuring the parameters of dipolar coupled spin systems from NMR spectra is currently a hard problem.  One approach, called pure frequency fitting, is to minimize by least squares the difference between the observable peak frequencies and the simulated transition frequencies \cite{CBB,Mahesh,Benzofuran,LAOCOON,LAOCOONOR,LEQUOR,MIMER,PANIC,PAREMUS}. The  major drawback of this approach is the requirement of spectral assignment, a manual procedure to determine which experimental peak corresponds to which simulated transition. To avoid spectral assignment, the straightforward strategy is to fit the spectrum, directly obtained from the thermal state via nonselective pulses, using a least squares algorithm \cite{TLS1,TLS2,DANSOM1,DANSOM2,DAISY1,WINDAISY1,BR}. This approach, called line shape fitting, is associated with immense computational resources and is seriously limited by the huge number of local minima. For this reason, evolutionary algorithms, which are able to search through many local minima, have been proposed and used for line shape fitting \cite{IC,GA,12s,GA2,GA3}, with impressive but still limited success. These methods are unable to cope with a large search space, making them suitable mainly for molecules with high symmetry and accurately known geometrical structure \cite{GA}. For both pure frequency fitting and line shape fitting methods, proper initial guess and bounds of the parameters are thus required to approach the desired solution. Additional spectra are necessary for this purpose, where Z-COSY \cite{Zcosy1, Zcosy2} and homonuclear decoupling \cite{Mrev81,Mrev82,Mrev83,Cory,Zhang} techniques are helpful for obtaining crucial clues to estimate certain parameters. Strategies based on multiple quantum coherence NMR \cite{highONMR,JCP07,GA} have been developed to reduce the number of local minima, exploiting the fact that the number of higher order transitions is much less than the number of single order transitions. 
The high order transitions can  be easily observed with 2D experiments. These experiments require the optimization of  the delay during the preparation period and therefore usually consumes long measuring times. Recently, theoretical strategies based on local control techniques were proposed through accessing the system partially  and an experimental demonstration was implemented in three spins with well known Hamiltonian using NMR \cite{local,local2,local3}.

In this article, we present NAFONS (Non-Assigned Frequency Optimization of NMR Spectra), a pure frequency fit program in which the spectral assignment problem is incorporated into a standard numerical optimization problem that can be addressed by a computer. Our global optimization strategy is based on the injection of random perturbations designed to enable the solver to escape local minima. The spectra to be fitted are obtained by standard 1D experiments. Experimentally NAFONS was applied to solve a 6-spin system with low symmetry, without prior knowledge of the interspin distances or order parameters and even without a first-order estimation of the parameters. The parameters of the Hamiltonian are well estimated in a few minutes and with no operator intervention. 

\section{Hamiltonian}
 Molecules dissolved in liquid crystal solvents usually present vanishingly small  intermolecular interactions and can be ignored. Nevertheless, the intramolecular dipolar couplings are present and scaled down by the order parameters. The Hamiltonian of the spin system can be represented as \cite{Levitt}
\begin{equation}
\mathscr{H} =   \sum_{j}{\mathscr{H}_{j}^{CS}+ \sum_{j,k>j}
\left({\mathscr{H}_{jk}^{DD}+ \mathscr{H}_{jk}^{J}}\right)}.
\label{Hamiltonian}
\end{equation}
%%%%%%%%%
Where
 \begin{eqnarray}
&&\mathscr{H}_{j}^{CS} =\pi \nu_{j}Z_{j}, \\
&&\mathscr{H}_{jk}^{DD} = \frac{\pi D_{jk}}{2}\times
 \begin{cases} 2Z_{j}Z_{k}, &\mbox{if heteronuclear,}\\
(2Z_{j}Z_{k}-X_{j}X_{k}-Y_{j}Y_{k}), & \mbox{if homonuclear,}
 \end{cases}\\
&&\mathscr{H}_{jk}^{J} =  \pi J_{jk}\times
 \begin{cases} Z_{j}Z_{k}, &\mbox{if heteronuclear,}\\
(Z_{j}Z_{k}+X_{j}X_{k}+Y_{j}Y_{k}), & \mbox{if homonuclear,}
 \end{cases}
\end{eqnarray}

\noindent
 $X_{i}$, $Y_{i}$, $Z_{i}$ denote the Pauli matrices with $i$ indicating the spin location, $\nu_{i}$ denotes the chemical shift of spin $i$,  $J_{ij}$ denotes the scalar coupling between spins $i$ and $j$, and $D_{ij}$ denotes the dipolar coupling. \\

%The dipolar couplings are much larger (up to 2-3 order of magnitudes) than the scalar couplings.  We thus firstly search for the $\{ \nu_{i},\, D_{ij}\}$, which are stored in a vector $\mathbf{x}$. We measure the scalar couplings in an isotropic solvent, {\it e.g.} chloroform, and then use them as initial guess in the anisotropic solvent to further adjust  all the parameters.

\section{NAFONS Algorithm}\label{Algorithm}

In NAFONS, the transition assignment is encoded in an objective function and each evaluation of this objective function will automatically determine a group of simulated transitions to those of the experimental group. Then, NAFONS choose for the individual transitions within the two groups. Spectral assignment can be modified at any step of the optimization of the parameters. 

In practice, the experimental frequencies of the $n$ peaks with largest integrals (where $n$
is a suitable and sufficiently large number) are extracted and stored  in a vector $\boldsymbol{F}^{exp}$.  The optimization procedure has  its start in an arbitrary Hamiltonian parameters vector, $\mathbf{x}_0$ and follows the next steps:
\begin{enumerate}
\setlength{\itemsep}{0pt}
\item  Find a minimizer $\mathbf{x}^*$ of $ f= \sum_{j}{({F}^{exp}_{j} - {F}^{sim}_{j})}^2$, where $\boldsymbol{F}^{sim}$ is the vector containing the frequencies, in increasing order, of the $n$ simulated transitions with largest integrals.
\item  Update the initial guess: $\mathbf{x}_0 \mapsto \mathbf{x}^*$.
\item Do 1 to 2 again, but with a perturbation $\boldsymbol{w}$ such that $ f= \sum_{j}{w_{j}({F}^{exp}_{j} - {F}^{sim}_{j})}^2$.
\item  Do 1 to 3 until the global minimum is reached.
\end{enumerate}
The way in which the objective function encodes the assignmet of the transitions considers that for each point of the parameter space, the group of simulated transitions is selected using their integrals and the assignment is done by sorting the frequencies in increasing order. This is a natural way of optimizing both the parameters and the assignment of the transitions. 

The problem to solve is represented as
\begin{equation}
\min_{\mathbf{x}\in  \Omega}\;\;f_{\boldsymbol{w}}(\mathbf{x}) = \sum_{j}{w_{j}({F}^{exp}_{j} - {F}^{sim}_{j})}^2, \label{problem}
\end{equation}
where $\mathbf{x}$ is the vector of parameters, $\Omega$ is the search domain, $\boldsymbol{F}^{exp}$
is the vector  of sorted experimental frequencies,
$\boldsymbol{F}^{sim}$ is the vector  of sorted simulated
frequencies and $\boldsymbol{w}$ is a vector of random weights. The
goal is to find a $\mathbf{x}^*$ that is a solution to problem
(\ref{problem}) for any value of  $\boldsymbol{w}$. In principle, this
is possible only for the optimal solution, in which case all of
the experimental and simulated peaks are in quasi-exact agreement, {\it i.e.} $\boldsymbol{F}^{exp} \simeq \boldsymbol{F}^{sim}$. In this case, we  have $(\boldsymbol{F}^{exp} - \boldsymbol{F}^{sim}) \simeq \mathbf{0}$, such that for all $ \boldsymbol{w} \in  \mathds{R}^{n}$, we have
$ \sum_{j}{w_{j}({F}^{exp}_{j} - {F}^{sim}_{j})}^2 \simeq  0$.

In a sense, this formulation of the problem is a way of avoiding suboptimal solutions by considering that the number of
objective functions that we could globally minimize to get the Hamiltonian is infinite. A  large number of objective functions which
do not share the same suboptimal solutions but that do share a same optimal solution were considered as was the fact that overlap, mainly for low
error suboptimal solutions, may ocurr. This approach also took into consideration that the problem is greatly overdetermined, due to the redundancy of single order quantum coherence spectra and that only a few elements of $\boldsymbol{w}$ are non-zero and still have a valid objective function $f_{\boldsymbol{w}}$. Finding a solution to (\ref{problem}) through the search of an  $\mathbf{x}^*$  that minimizes $f(\mathbf{x}) = \sum_{j}{({F}^{exp}_{j} - {F}^{sim}_{j})}^2$ is the fist step.

The second step includes the use of this $\mathbf{x}^*$  as the initial guess for minimizing a randomly modified objective function
of the form $f_{\boldsymbol{w}}(\mathbf{x}) = \sum_{j}{w_{j}({F}^{exp}_{j} - {F}^{sim}_{j})}^2$, where the elements of $\boldsymbol{w}$
are chosen randomly to be either 0 or 1. If the solution $\mathbf{x}^*$ is a global minimizer of $f$, then the solver will not modify the solution,
otherwise the solver continues the optimization with the modified objective function $f_{\boldsymbol{w}}$. 

These last two steps can be done repeatedly in a $M$ times loop, where the loop has the form shown in the next equation
\begin{equation}
\begin{split}
\mbox{Solve}\;\;\;\;\; &\min_{\mathbf{x}\in  \Omega}\;\;f(\mathbf{x}) = \sum_{j}{({F}^{exp}_{j} - {F}^{sim}_{j})}^2, \\
\mbox{Solve}\times M \;\;\;\;\; &\min_{\mathbf{x}\in  \Omega}\;\;f_{\boldsymbol{w}}(\mathbf{x}) = \sum_{j}{w_{j}({F}^{exp}_{j} - {F}^{sim}_{j})}^2.
\end{split}
\label{loop}
 \end{equation}
The equilibrium state of this process is the commonly shared optimal solution.

The final step considers the use of an interior-point approach \cite{IP1} and the inclusion of pattern search \cite{PS1} to locate the optimum. Once the solution is found, a least squares fit \cite{LS1} of the spectrum line shape is finally done to adjust the decoherence rates of each spin and the scalar couplings.  As an altenative, a random walk approach in which the loop (\ref{loop}) was replaced by a single step taken into the direction that minimizes $f$ followed by an other single step in the direction that minimizes a randomly chosen $f_{\boldsymbol{w}}$ can also be used.

\section{Experiment Results}\label{Results and Discussion}
A Bruker 600 MHz Avance spectrometer with a 5mm dual $^{1}H - ^{19}F$ probe was used to analyze the 6 spin system ($2$ fluorine and $4$ protons) of the 2,3 difluorobenzaldehide (23DFBA, C$_{7}$H$_{4}$F$_{2}$O), figure \ref{2,3-DFBA}, dissolved in the liquid crystal ZLI-1132. The temperature was controlled at $284$K.  The full  internal Hamiltonian was obtained through the fitting different spectra: {\bf 1)} fluorine spectrum with proton decoupling, {\bf 2)} proton spectrum with fluorine decoupling, {\bf 3)} spectra obtained by selective transition pulses based on spectrum 2), {\bf 4)} fluorine spectrum without proton decoupling, and {\bf 5)} proton spectrum without fluorine decoupling. Standard composite decoupling pulses, {\it i.e.}  GARP \cite{GARP} were used to decouple fluorine spins and SPINAL-64  \cite{SPINAL} to decouple proton nuclei. The selective transition pulses were Guassian shaped pulses with duration of $20$ ms.

 NAFONS approach was tested in a highly simplified configuration: the search was done directly on the chemical shifts and dipolar couplings (without assuming or guessing the molecular geometry and order parameters), the optimization was done without a proper initial guess (0 Hz for each parameter) and without proper bounds ($\pm$2500 Hz for each parameter), the diagonalizations of the Hamiltonian were done with a general QZ algorithm \cite{QZ} and the program was implemented in MATLAB.

The full internal Hamiltonian parameters (chemical shifts, $J$ and $D$ couplings) are listed in Table \ref{FullTable}. The errors, shown in parenthesis, were estimated by comparing the values obtained from the different fitted spectra and by using the standard deviation considering  gaussian noise.

\subsection{Fluorine Spectra with Proton Decoupling}
\begin{figure}
\centering
\includegraphics[width=6in]{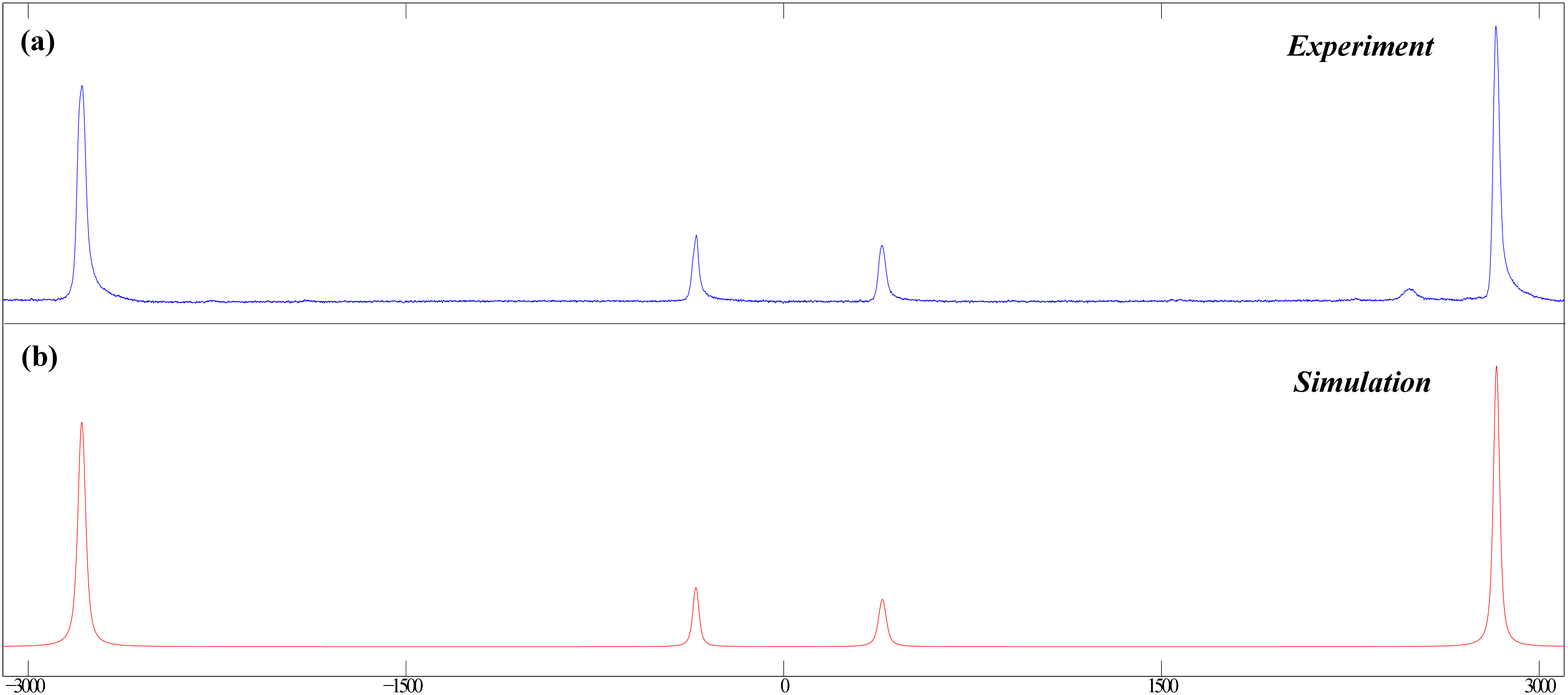}
\caption[fu]
  {Fluorine spectra with proton decoupling,
  obtained in experiment (a) and by simulation (b). The agreement indicates a reliable estimation of the parameters.}
  \label{fu}
\end{figure}

The fluorine spectrum is shown in figure \ref{fu}, where the four transitions are present and used for optimization. Convergence was reached within a second. The agreement between the simulation and the experiment present in Figure \ref{fu}b, indicates a reliable estimation of the parameters. An extra ``junk" peak in the experimental spectrum at $\sim 2500$Hz was possibly due to the imperfection of the decoupling. The results for the chemical shifts (in Hz) of F$_5$ and F$_6$ are: $-885(3)$  and $948(2)$ with respect to a transmitter frequency. The result for the dipolar coupling (in Hz) is: $-1589(7)$.

 The chemical shifts (up to a scaling factor) were further verified by a 2D experiment using Lee Goldburg decoupling technique \cite{LG1,LG2}. In this technique a radio frequency off resonance is applied according to $\Delta LG = \frac{\sqrt{2}}{2} \omega_{1}$ causing an effective magnetic field in the rotating frame inclined at the magic angle (in relation to the static magnetic field) $\theta = tan^{-1}(\sqrt{2})$ and therefore, the dipolar coupling is refocused and only the chemical shift evolves during $t_{1}$. The values obtained using this technique were $933.98$ and $-871.78$ Hz for F(5) and F(6) respectively. The dipolar coupling obtained was $-1576.49$Hz.

\subsection{Proton Spectra with Fluorine Decoupling}
\begin{figure}
\centering
\includegraphics[width=6in]{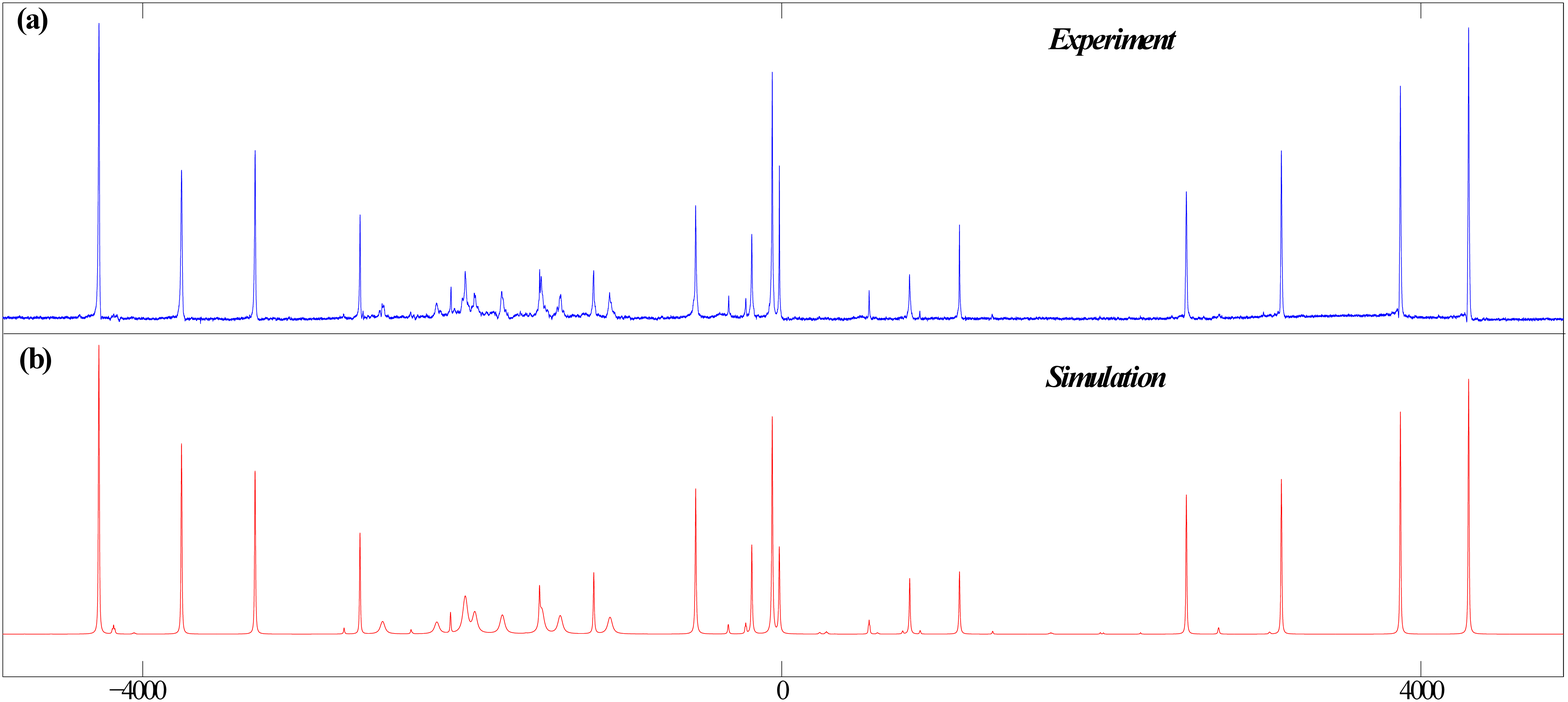}
\caption[pu]
  {Proton spectra with fluorine decoupling,
  obtained in experiment (a) and by simulation (b). The agreement indicates a reliable estimation of the parameters.}
  \label{pu}
\end{figure}

The 1D proton spectrum of 23DFBA is shown in figure \ref{pu}. The 26 transitions with largest integrals were selected for optimization. For several trials, convergence was usually reached within 10 minutes. The mean frequency error was 0.25 Hz, probably due to line-overlap, which was not taken into consideration. The chemical shifts (in Hz) for H$_1$, H$_2$, H$_3$ and H$_4$ are  $-1770(3)$, $-149(2)$, $172(2)$ and $-234(3)$ respectively. The agreement between the simulation and the experiment  is present in figure \ref{pu}b and indicates a reliable estimation of the couplings shown in Table \ref{FullTable}.  Some small differences in the relative heights of the transitions are present and might be explained as an imperfection in our way of modelling decoherence.

 Around -2000 Hz, the cluster of transitions with strong decoherence corresponds to H$_1$. These transitions are closely distributed around its
chemical shift value ($-1770$Hz) whose dipolar coupling involving H$_1$ are relatively small ($<$ 450 Hz). The two sets of 4 transitions with
high amplitudes on the extreme left and extreme right of the spectrum  both correspond to a mix of H$_2$ and H$_3$
transitions. They are at the extremes of the spectrum due to the large coupling (-2166 Hz) between H$_2$ and H$_3$. The
transitions corresponding to H$_4$ are distributed on a width of $\sim 1600$ Hz around the centre of the spectrum. This is mainly
due to the coupling between H$_3$ and H$_4$ (-931 Hz). In this case the experimental chemical shifts (up to a scaling factor) obtained by a 2D
experiment using Lee Goldburg gave results quite different from those expected, H$1$ = $-1104.7$,H$2$ = $505.8$, H$3$ = $829.1$ and H$4$ = $426.4$. Nevertheless, the results from table \ref{FullTable} are consistent with the experimental data and were used in further experiments.
 
\subsection{Measuring Dipolar Couplings Between Heteronuclear Spins}\label{SectheteroDipolar} 
\begin{figure}
\centering
\includegraphics[width=6in]{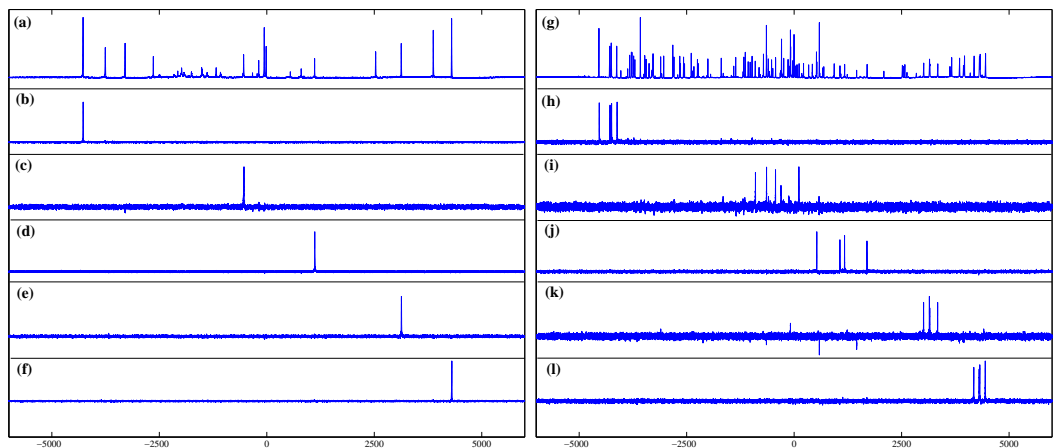}
\caption[fc]
{Spectra for extracting the dipolar couplings between heteronuclei. Full proton spectrum with fluorine decoupling (a) and corresponding subspectra obtained by transition selective pulses (b)-(f). Full proton spectrum without fluorine decoupling (g) and corresponding subspectra obtained by the same transition selective pulses (h)-(l).} 
 \label{Tslec}
\end{figure}

With the proton chemical shifts and proton homonuclear dipolar couplings sumarized in Table
\ref{FullTable}, the proton Hamiltonian can be diagonalized. Each eigenvector can be expressed in the computational basis,
{\it i.e.} $\{|0\rangle$, $|1\rangle\}$ and a map between transitions and energy levels can be built. There are about 10 well resolved
peaks in Figure \ref{pu}(a) that can be excited individually with transition selective pulses (Gaussian shaped, 20 ms). Five experimental spectra
obtained through transition selective pulses of certain lines are shown  in Figures \ref{Tslec}(b)-(f). Figure \ref{Tslec}(a) is the 1D proton decoupled from fluorine formerly used in  figure \ref{pu}(a)  and is at the top of Figure \ref{Tslec} as a reference for figures \ref{Tslec}(b)-(f). Figure \ref{Tslec}(g) shows the full proton spectrum without fluorine decoupling.

Each of these 5 transitions corresponds to a density matrix, that can be written as the external product of the
eigenstates involved in the transition, represented as
\begin{equation}\label{denMtrans}
 \rho_{ij}^{H} = |E_{i}\rangle\langle E_{j}|,
\end{equation}
where $|E_{i}\rangle$ denotes one eigenstate of the proton Hamiltonian. Then, the decoupling channel is switched off  so a proton coupled to 
fluorine spectrum is acquired and the heteronuclear coupling is evident. The resulting spectra is shown in Figures  \ref{Tslec}(h)-(l) in the same order as 
in Figures \ref{Tslec}(b)-(f) for comparison. The corresponding states are 
 \begin{equation}\label{denMtrans}
 \rho_{ij} = \rho_{ij}^{H}\otimes I,
\end{equation}
where $I$ denotes a 4$\times$4 identity matrix, representing the state of the two fluorine spins. We use the $\{\rho_{ij}\}$ as the input
states to simultaneously analyze  the spectra shown in Figures \ref{Tslec}(h)-(l) and extract all the heteronuclear dipolar
couplings. The chemical shifts are allowed to vary $\pm\,50\,\mbox{Hz}$ from their values obtained with decoupling pulses. For several trials, convergence is usually reached within 10 minutes. 

\subsection{Complete Fluorine and Proton Spectra}
\begin{figure}
\centering
\includegraphics[width=6in]{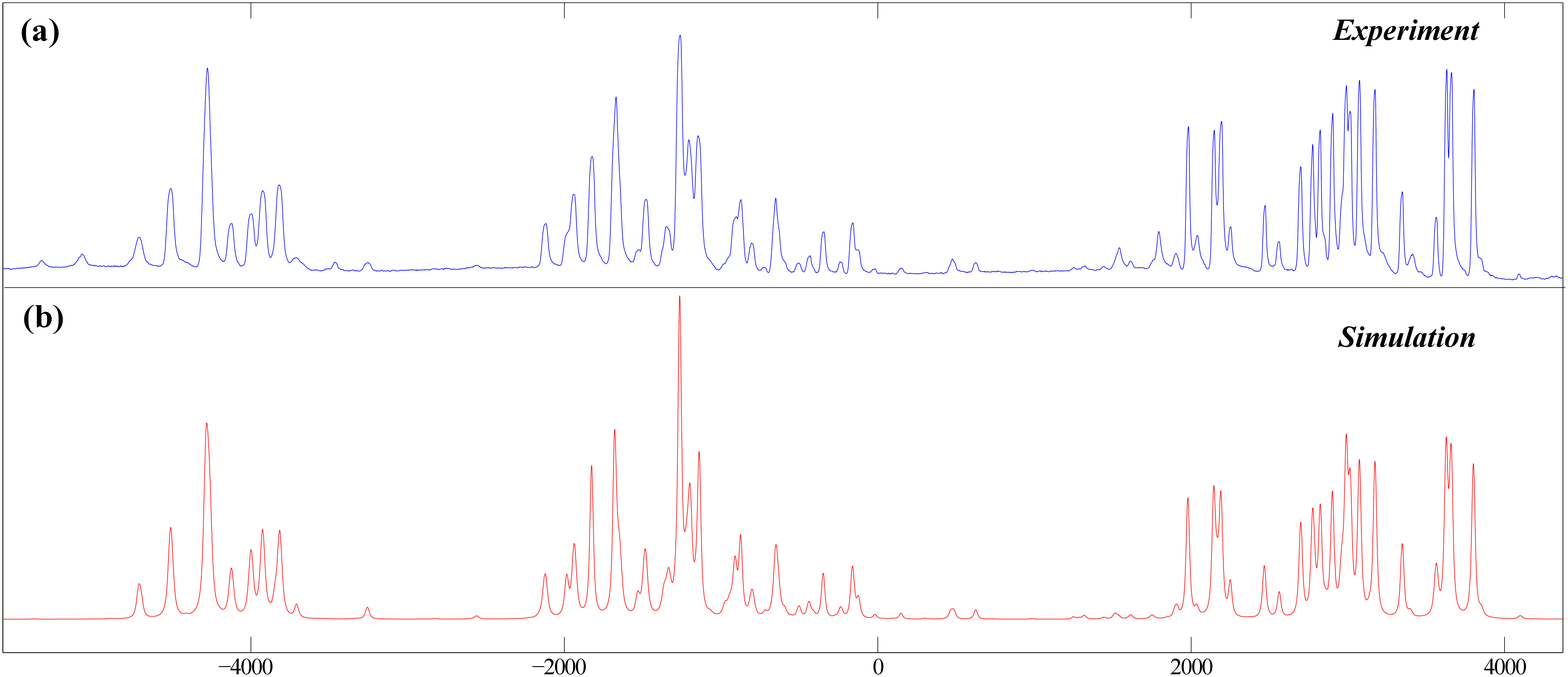}
\caption[fc]
  {Fluorine spectra without proton decoupling,
  obtained in experiment (a) and by simulation (b). The occasional difference in heights is probably due to our modelling of decoherence (see text).}
  \label{fc}
\end{figure}

\begin{figure}
\centering
\includegraphics[width=6in ]{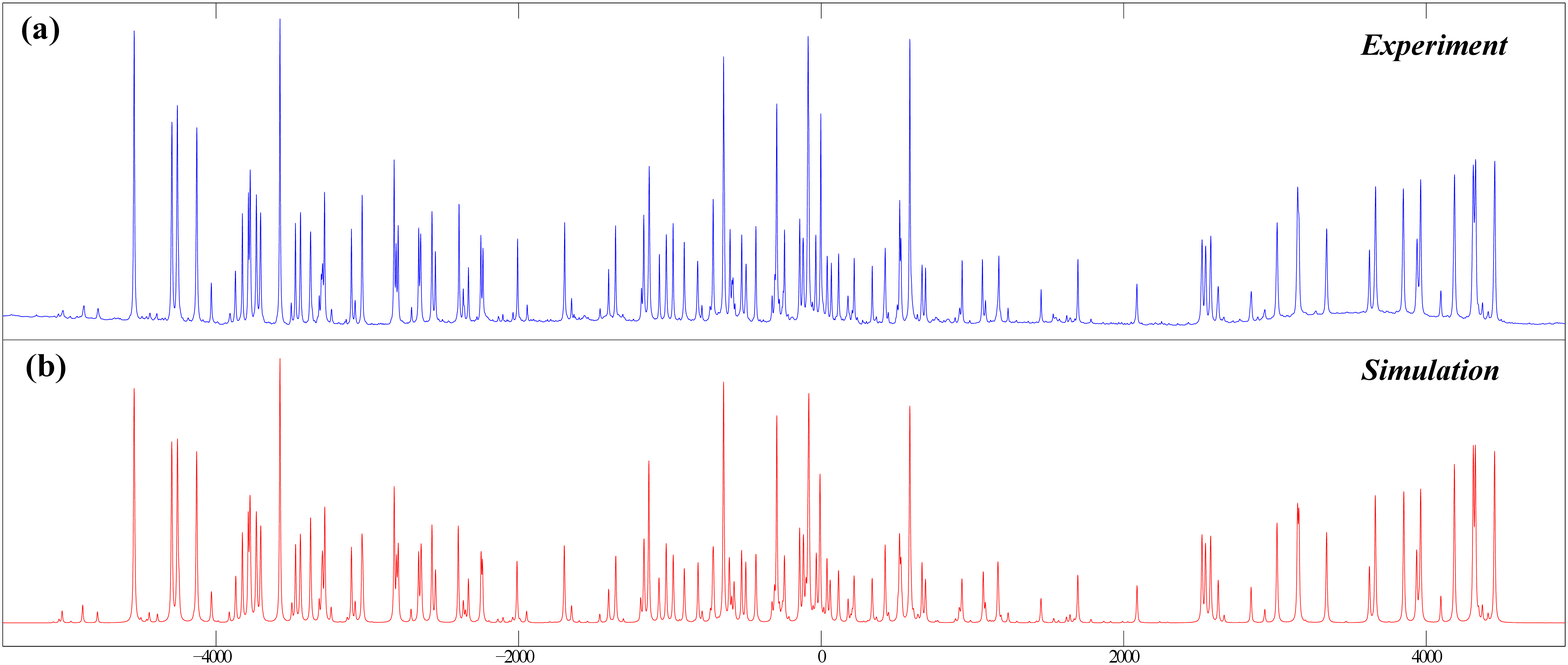}
\caption[pc]
  {Proton spectra without fluorine decoupling,
  obtained in experiment (a) and by simulation (b). The occasional difference in heights is probably due to our modelling of decoherence (see text).}
  \label{pc}
\end{figure}

Once the full internal Hamiltonian is obtained, these parameters were used to fit the complete fluorine and proton spectra using least squares on the spectral line shape, mainly to adjust the decoherence rate of the spins and the scalar couplings.  During this fit, the Hamiltonian parameters changed less than 1\% from the values previously obtained. These changes were probably due to line-overlap, which was not taken into account during the pure frequency fit. These results for the fluorine and proton spectra are shown in Figures \ref{fc} and  \ref{pc} respectively and the parameters are listed in Table \ref{FullTable}. The ${T_{2}}^*$ (in ms) for H$_1$, H$_2$, H$_3$, H$_4$, F$_5$ and F$_6$ are respectively: $80.2(0.3)$, $65.8(0.2)$, $60.4(0.3)$ $62.4(0.2)$, $11.6(0.3)$ and $15.9(0.1)$. 

The selection of specific transitions helps  to minimize the complexity of the spectrum with strong coupling as in molecules dissolved in liquid crystals. Not only the number of transitions is much less, but the number of possible assignments for these remaining transitions is also reduced. For these experiments, only 21 transitions are used to extract the heteronuclear dipolar couplings  and 4!$\cdot$5!$\cdot$4!$\cdot$4!$\cdot$4! assignments were possible instead of 21!. This experiment also made possible to identify which transitions should be used for the analysis, otherwise, the presence of overlap becomes an obstacle to this step.

\section{Discussion}

In the approach introduced by Castellano and Bothner-By \cite{CBB}, the differences between the observable peak frequencies and the simulated transition frequencies are minimized using a least squares algorithm. The well-known major drawback of this method is the requirement of spectral assignment, to establish which experimental peak correspond to which simulated transition. In traditional programs such as LAOCOONOR \cite{LAOCOONOR}, PANIC \cite{PANIC} and LEQUOR \cite{LEQUOR}, both the parameters and the spectral assignment  have to be adjusted by the operator before each trial fitting. Successful attempts of automating the assignment procedure have been reported in programs such as PAREMUS \cite{PAREMUS} and MIMER \cite{MIMER}, but these are limited to simple solutes in isotropic solvents. Thus, in traditional pure frequency fitting algorithms, the procedure of spectral assignment is still the most decisive and difficult step, rapidly rendering them impossible to apply, especially when the molecular geometry and orientational parameters are unknown or difficult to guess.

Automatic methods which do not require spectral assignment have been developed as an alternative. These approaches, called integral transform (IT) and total line shape (TLS), use the full spectral line shape. In the IT approach, introduced by Diehl, S\'{y}kora and Vogt \cite{IT1}, the spectrum  is transformed into a small set of coefficients by means of linear integral transforms using orthogonal bases. The differences between the coefficients obtained from the experimental spectrum and those obtained from the simulated one are minimized with a standard optimization routine. In the TLS approach, the total line shape of the NMR spectrum is fitted. The idea was first demonstrated by Glidewell, Rankin and Sheldrick \cite{TLS1}, and also studied by Heinzer \cite{TLS2}. A matrix method derived from a general formulation of the least squares problem was then developed by Stephenson and Binsch \cite{DANSOM1,DANSOM2}. The originality in their method was the use of cross-correlation functions to smooth the landscape, other techniques such as spectrum broadening \cite{BR} and integral curves \cite{IC} have also been proposed for this purpose. This method, and its subsequent modifications - DAISY \cite{DAISY1} and WIN-DAISY \cite{WINDAISY1}, were later improved by the use of Genetic Algorithms (GA's) \cite{IC,GA,GA2,GA3,12s}, which are able to search through many basin of attractions. It is known that for GA's, when the search ranges become too large, there is insufficient coverage of the parameter space to locate the global minimum \cite{GA}. Some improvements can be obtained by the use of Evolutionary Strategies (ES's), which usually converge faster than GA's \cite{GA}. Evolutionary algorithms such as GA's and ES's are thus suitable only for molecules with high symmetry and with accurately known geometrical structure \cite{GA}. In general, both the IT and TLS approaches suffer from severe limitations: they are computationally  much slower than frequency fitting \cite{IT2,IT3,PAREMUS}; their global optimization strategy is either absent or operational only in small search spaces; the operator has hardly any means of interacting with the program to increase its efficiency. Due to these limitations, automatic analysis is not routinely employed \cite{JCP07}, and the Castellano-Bothner-By approach is still by far the most widely used \cite{LAOCOONOR}, despite the requirement of spectral assignment. 

The originality of our approach can now be seen: it is  a pure frequency fit program which  incorporates the spectral assignment problem into a standard numerical optimization problem that can be addressed by a computer. In contrast with traditional automatic methods, evaluation of the objective function does not require the expensive computation of the spectral line shape. Moreover, our global optimization strategy, based on the injection of randomness, is able to cover a large search space without getting trapped in local minima. The most interesting feature of our approach is perhaps its compatibility with operator interventions. In fact, at any moment, the operator could pause the program, so as to visually compare the spectra and possibly choose to impose constraints on the spectral assignment, gradually removing the suboptimal attractors from the landscape. NAFONS new approach successfully addresses the fundamental problems usually encountered in spectral analysis. 

\section{Conclusion and Outlooks}\label{Conclusion}
A new method for solving NMR spectra of solutes dissolved in liquid-crystals is shown as well as an experiemental application to solve  a 6-spin
system with very low symmetry. This was done without prior knowledge or assumptions on the interspin distances or order parameters, which contrasts with previous results in \cite{WINDAISY1,DANSOM1,DANSOM2,LAOCOONOR,LAOCOON,Mahesh,Benzofuran,structure,IC,GA,JCP07,12s,GA2,GA3}. This method includes a new spectral analysis program - NAFONS, and experimental techniques to simplify spectral analysis for extracting the dipolar couplings between heteronuclear spins. In contrast with traditional pure frequency fitting methods \cite{CBB,Mahesh,Benzofuran,LAOCOON,LAOCOONOR,LEQUOR,PANIC,MIMER,PAREMUS}, NAFONS does not require spectral assignment and is  fully automatic. As for line shape fitting methods \cite{TLS1,TLS2,BR,IC,GA,DANSOM1,DANSOM2,DAISY1,WINDAISY1,12s,GA2,GA3}, evaluation of an objective function does not involve the expensive computation of the spectral line shape and the global optimization strategy can cope with a large search space. These results should be helpful to implement spectral analysis of dipolar coupled systems and can be extended to larger systems. Using these methods, it is now possible to create a library of molecules for chemical structure analysis that can be used as well in other fields such as Quantum Information Processing and Quantum Computing.

\section{Acknowledgments}
The authors acknowledge professor D. G. Cory, professor D. Suter, Dr. O. Moussa and Dr. D. Burgarth for helpful discussions.\\

NAFONS program can be obtained upon request at trottier.denis.alexandre@gmail.com.

\newpage

\begin{table}
\begin{tabular}{|c|c|c|c|c|c|c|}
\hline  &   H$_1$ &  H$_2$ &  H$_3$ &  H$_4$ & F$_5$ & F$_6$ \\
\hline H$_1$ &  -1770(3)  & -424(3)   &  -144(3)  &  -154(2) &  -1505(4)  & -232(3)   \\
\hline H$_2$ &  0.38   & -149(2)   & -2166(8)   & -368(4)  & -42(4)   &  -106(2) \\
\hline H$_3$ & -0.05   & 7.88    & 172(2)   & -931(5) & -62(4) &  -46(3) \\
\hline H$_4$ &  0.36   & 1.75   & 7.70   & -234(3) & -236(3)  & -384(3)    \\
\hline F$_5$ &  -0.04   & 5.56  &  1.43  & 8.14  & -885(3) & -1589(7)  \\
\hline F$_6$ &  -0.73   &  1.45  &  4.35  &9.82  & 20.75   &  948(2)  \\
\hline
\end{tabular}
 \caption {Full internal Hamiltonian of  2,3-Difluorobenzaldehyde measured in the liquid crystal solvent ZLI-1132.  The chemical shifts (in Hz) are shown in the diagonal and are with respect to transmitter frequency at 600.13Hz and 564.62 MHz for proton and fluorine spins respectively. The scalar couplings ($J$) are shown in the lower part of the diagonal and were obtained by the conventional 2D NMR experiments, e.g. J-Resolved, COSY, etc. and the dipolar couplings ($D$) are in the higher part of the diagonal and were obtained through the NAFONS method. The numbers in parenthesis are the errors obtained. }  \label{FullTable}
\end{table}

\begin{figure}
\includegraphics[width=2in]{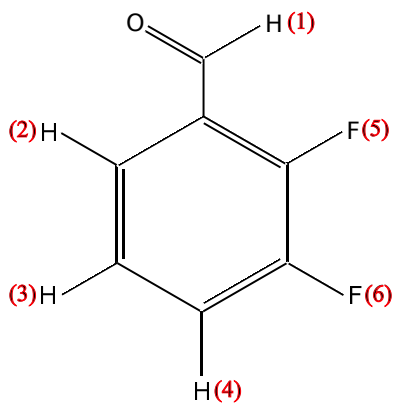} 
\caption{Molecular structure of 2,3-Difluorobenzaldehyde and the
spin labelling.} \label{2,3-DFBA}
\end{figure}

\end{document}